\documentclass[12pt,thmsa]{article}
\usepackage{amssymb}
\usepackage{amsmath}
\usepackage{graphicx}

\def\R{\mathbb{R}}
\def\C{\mathbb{C}}
\def\N{\mathbb{N}}

\def\Z{\mathbb{Z}}

\newtheorem{theorem}{Theorem}

\newtheorem{lemma}[theorem]{Lemma}
\newtheorem{proposition}[theorem]{Proposition}
\newtheorem{definition}[theorem]{Definition}
\newtheorem{example}[theorem]{Example}
\newtheorem{remark}[theorem]{Remark}

\begin{document}

\author{\textbf{Yuri G.~Kondratiev\thanks{Support by FCT, POCTI - Programa 
Operacional Ci\^encia, Tecnologia, Inova{\c c}\~ao, FEDER}} \\
Faculty of Mathematics, University of Bielefeld, D-33615 Bielefeld\\
BiBoS, University of Bielefeld, D-33615 Bielefeld\\
Institute of Mathematics, NASU, U-252601 Kiev\\
kondrat@mathematik.uni-bielefeld.de \and
\textbf{Maria Jo\~{a}o Oliveira}$^*$ \\
Univ.~Aberta, P-1269-001 Lisbon\\
GFMUL, University of Lisbon, P-1649-003 Lisbon\\
BiBoS, University of Bielefeld, D-33615 Bielefeld\\
oliveira@cii.fc.ul.pt}

\title{Invariant measures for Glauber dynamics of continuous systems}
\date{}
\maketitle

\begin{abstract}
We consider Glauber-type stochastic dynamics of continuous systems 
\cite{BCC02}, \cite{LK03}, a particular case of spatial birth-and-death 
processes. The dynamics is defined by a Markov generator in such a way that 
Gibbs measures of Ruelle type are symmetrizing, and hence invariant for the 
stochastic dynamics. In this work we show that the converse statement is also 
true. Namely, all invariant measures satisfying Ruelle bound condition are 
grand canonical Gibbsian for the potential defining the dynamics. The proof 
is based on the observation that the well-known Kirkwood-Salsburg equation 
for correlation functions is indeed an equilibrium equation for the 
stochastic dynamics.
\end{abstract}

\textbf{MSC Classification:} 37L40, 60J75, 60J80, 60K35, 82C21, 82C22
\newpage 

\section{Introduction}

For Gibbs states $\mu$ on the space $\Gamma$ of all locally finite subsets 
(configurations) of $\R^d$ and being either of the Ruelle type or 
corresponding to a positive potential, it has been constructed in the recent 
work \cite{LK03} an equilibrium Glauber-type dynamics on $\Gamma$ having $\mu$ 
as an invariant measure. That is, $H^*\mu=0$ where $H^*$ is the dual operator 
of the generator $H$ of the dynamics. The dual 
relation between observables and states yields a further interpretation for 
this invariance result, namely, the Gibbs states as above are stationary 
distributions of the Glauber dynamics generated by $H$. In this work we 
study the converse problem related to the question of whether all invariant 
measures are Gibbsian. For this purpose, we begin by enlarging the class 
of invariant measures to a new class of measures outside of the semigroup 
setting. These new elements, called infinitesimally invariant measures (corresponding to 
$H$), are probability measures $\mu$ on $\Gamma$ with finite moments of all 
orders which satisfy Ruelle bound condition, and
$$
\int_\Gamma (HF)(\gamma)\,d\mu(\gamma)=0
$$
for all functions $F$ in a proper dense set in the space $L^1(\Gamma,\mu)$.
We effectively formulate the notion of infinitesimally invariant measures by using the 
combinatorial harmonic analysis on configuration spaces introduced and 
developed in \cite{KoKu99}, \cite{KoKu99c}, \cite{K00} (Section 
\ref{Section1}). The special 
nature of this analysis yields, in particular, natural relations between 
states, observables, and correlation measures. We exploit these relations 
to show that for potentials fulfilling the usual integrability, 
stability, and lower regularity conditions, any infinitesimally invariant measure is 
Gibbsian (Theorem \ref{Prop5}). In particular, this result applies to 
any invariant measure of the stochastic dynamics generated by $H$. 
This answers an old open question usually known as the Gibbs conjecture 
for stochastic dynamics. Originally this question was formulated for the 
Hamiltonian case, and then generalized to other dynamics. For the 
Hamiltonian case the problem is partially solved, and the most meaningful 
contributions obtained on this direction are essentially due to 
B.~M.~Gurevich, Ya.~G.~Sinai, Yu.~M.~Suhov (see e.g.~the review work 
\cite{Do94} and the references therein). Considerable progress in the 
stochastic dynamics direction have been achieved in \cite{HS81} and 
\cite{Fr82}, \cite{Fr86}, \cite{FLO97}, \cite{FRZ98} for the diffusion 
dynamics of infinite lattice systems over $\Z^d$, for $d\leq 2$. Recently, 
these results have been generalized in \cite{BRW02} to any dimension and, 
moreover, to spin spaces not necessarily compact. In our case, we obtain an 
analogous result for Glauber-type stochastic dynamics of continuous systems. 
Besides the statement formulated in 
Theorem \ref{Prop5}, the proof itself encloses an additional interpretation 
for the well-known Kirkwood-Salsburg equation for correlation functions. More 
precisely, it shows that the Kirkwood-Salsburg equation is indeed an 
equilibrium equation of the stochastic dynamics generated by $H$. As an aside, the existence of 
solutions for this equation, in the case of positive potentials in the high 
temperature-low activity regime, can easily be demonstrated. We postpone 
this subject to a forthcoming publication devoted solely to the problem of 
existence of non-equilibrium Glauber dynamics corresponding to more general 
potentials.   

\section{Harmonic analysis on configuration spaces\label{Section1}}

The configuration space $\Gamma := \Gamma_{\R^d}$ over $\R^d$, $d\in \N$, is 
defined as the set of all locally finite subsets of $\R^d$,
\[
\Gamma :=\left\{ \gamma \subset \R^d:\left| \gamma_\Lambda\right| <\infty 
\hbox{
for every compact }\Lambda\subset \R^d\right\} , 
\]
where $\vert \cdot \vert$ denotes the cardinality of a set and 
$\gamma_\Lambda := \gamma \cap \Lambda$. As usual we identify each 
$\gamma \in \Gamma$ with the non-negative Radon measure 
$\sum_{x\in \gamma}\varepsilon_x\in \mathcal{M} (\R^d)$, where 
$\varepsilon_x$ is the Dirac measure with mass at $x$, 
$\sum_{x\in \emptyset}\varepsilon_x$ is, by definition, the zero measure, and 
$\mathcal{M} (\R^d)$ denotes the space of all non-negative Radon measures on 
the Borel $\sigma$-algebra $\mathcal{B}(\R^d)$. This procedure allows to 
endow $\Gamma$ with the topology induced by the vague topology on 
$\mathcal{M} (\R^d)$. We denote the Borel $\sigma$-algebra on $\Gamma$ by 
$\mathcal{B}(\Gamma)$.

Let us now consider the space of finite configurations 
\[
\Gamma_0 := \bigsqcup_{n=0}^\infty \Gamma^{(n)},
\]
where $\Gamma^{(n)} := \Gamma^{(n)}_{\R^d} := \{ \gamma\in \Gamma: \vert \gamma\vert = n\}$ for $n\in \N$ and $\Gamma^{(0)} := \{\emptyset\}$. For $n\in \N$, there is a natural bijection between the space $\Gamma^{(n)}$ and the 
symmetrization $\widetilde{(\R^d)^n}\diagup S_n$ of the set 
$\widetilde{(\R^d)^n}:= \{(x_1,...,x_n)\in (\R^d)^n: x_i\not= x_j \hbox{ if } i\not= j\}$ under the permutation group $S_n$ over $\{1,...,n\}$ acting on 
$\widetilde{(\R^d)^n}$ by permuting the coordinate index. This 
bijection induces a metrizable topology on $\Gamma^{(n)}$ and we will 
endow $\Gamma_0$ with the topology of disjoint union of topological spaces. 
By $\mathcal{B}(\Gamma^{(n)})$ and $\mathcal{B}(\Gamma_0)$ we denote the 
corresponding Borel $\sigma$-algebras on $\Gamma^{(n)}$ and $\Gamma_0$, 
respectively.   

We proceed to consider the $K$-transform. Let $\mathcal{O}_c(\R^d)$ denote 
the set of all compact sets in $\R^d$, and for any 
$\Lambda\in \mathcal{O}_c(\R^d)$ let 
$\Gamma_\Lambda := \{\eta\in \Gamma: \eta\subset \Lambda\}$. Evidently 
$\Gamma_\Lambda = \bigsqcup_{n=0}^\infty \Gamma_\Lambda^{(n)}$, where
$\Gamma_\Lambda^{(n)}:= \Gamma_\Lambda \cap \Gamma^{(n)}$ for all 
$n\in \N_0$, leading to a situation similar to the one for $\Gamma_0$, 
described above. We endow $\Gamma_\Lambda$ with the topology of the disjoint 
union of topological spaces and with the corresponding Borel $\sigma$-algebra 
$\mathcal{B}(\Gamma_\Lambda)$. To define the $K$-transform let us 
consider the space $B_{bs}(\Gamma_0)$ of all complex-valued bounded
$\mathcal{B}(\Gamma_0)$-measurable functions $G$ with bounded support, i.e., 
$G\!\!\upharpoonright _{\Gamma _0\backslash 
\left(\bigsqcup_{n=0}^N\Gamma _\Lambda ^{(n)}\right) }\equiv 0$ for some 
$N\in\N_0, \Lambda \in \mathcal{O}_c(\R^d)$. The $K$-transform of any 
$G\in B_{bs}(\Gamma_0)$ is the mapping $KG:\Gamma\to\C$ defined at each 
$\gamma\in\Gamma$ by
\begin{equation}
(KG)(\gamma ):=\sum_{{\eta \subset \gamma}\atop{\vert\eta\vert < \infty} }
G(\eta ).
\label{Eq2.9}
\end{equation}
Note that for every $G\in B_{bs}(\Gamma_0)$ the sum in (\ref{Eq2.9}) has 
only a finite number of summands different from zero and thus $KG$ is a 
well-defined function on $\Gamma$. Moreover, if $G$ has support described as 
before, then the restriction $(KG)\!\!\upharpoonright _{\Gamma _\Lambda }$ is a
$\mathcal{B}(\Gamma_\Lambda)$-measurable function and 
$(KG)(\gamma) 
= (KG)\!\!\upharpoonright _{\Gamma _\Lambda }\!\!(\gamma_\Lambda)$ 
for all $\gamma\in\Gamma$, i.e., $KG$ is a cylinder function. In addition, 
for any $L\geq 0$ such that $\vert G\vert\leq L$, one finds
$\vert (KG)(\gamma)\vert\leq L(1+\vert\gamma_\Lambda\vert)^N$ for all 
$\gamma\in\Gamma$, i.e., $KG$ is polynomially bounded. It has been shown in 
\cite{KoKu99} that the $K$-transform is indeed a linear isomorphism between 
the spaces $B_{bs}(\Gamma_0)$ and 
$\mathcal{FP}_{bc}(\Gamma):= K\left(B_{bs}(\Gamma_0)\right)$. This leads, in 
particular, to an explicit description of all functions in 
$\mathcal{FP}_{bc}(\Gamma)$ which may be found in \cite{KoKu99} and 
\cite{KoKuOl00b}. However, throughout this work we shall only make use of the 
above described cylindricity and polynomial boundeness properties fulfilled 
by the elements in $\mathcal{FP}_{bc}(\Gamma)$. The inverse mapping of the 
$K$-transform is defined on $\mathcal{FP}_{bc}(\Gamma)$ by 
\[
\left( K^{-1}F\right) (\eta ):=\sum_{\xi \subset \eta }(-1)^{|\eta
\backslash \xi |}F(\xi ),\quad \eta \in \Gamma _0. 
\]

Besides the functions in $B_{bs}(\Gamma_0)$ we also consider the so-called 
coherent states $e_\lambda(f)$ of $\mathcal{B}(\R^d)$-measurable functions 
$f$, defined by
\[
e_\lambda (f,\eta ):=\prod_{x\in \eta }f\left( x\right) ,\ \eta \in
\Gamma _0\!\setminus\!\{\emptyset\},\quad  e_\lambda (f,\emptyset ):=1.
\]
For a $\mathcal{B}(\R^d)$-measurable function $f$ with compact support, we  
observe that the image of $e_\lambda(f)$ under the $K$-transform is still a 
well-defined function on $\Gamma$ and has an especially simple form given by
\[
\left( Ke_\lambda (f)\right) (\gamma )=\prod_{x\in \gamma }(1+f(x)),\quad 
\gamma\in \Gamma. 
\]

From the algebraic point of view let us consider the $\star$-convolution 
defined on $\mathcal{B}(\Gamma_0)$-measurable functions $G_1$ and $G_2$ by 
\[
(G_1\star G_2)(\eta ):=\sum_{(\eta _1,\eta _2,\eta _3)\in \mathcal{P}_3(\eta
)}G_1(\eta _1\cup \eta _2)G_2(\eta _2\cup \eta _3),\quad \eta \in \Gamma _0, 
\]
where $\mathcal{P}_3(\eta )$ denotes the set of all partitions of $\eta $ in
three parts which may be empty \cite{KoKu99}. It is straighforward to verify 
that the space of all $\mathcal{B}(\Gamma_0)$-measurable functions endowed 
with this product has the structure of a commutative algebra with unit element 
$e_\lambda(0)$. Furthermore, for every $G_1, G_2\in B_{bs}(\Gamma_0)$ we have 
$G_1\star G_2\in B_{bs}(\Gamma_0)$, and
\begin{equation}
K\left( G_1\star G_2\right) =\left( KG_1\right) \cdot \left( KG_2\right)
\label{1.5}
\end{equation}
cf.~\cite{KoKu99}. The $\star$-convolution applied, in particular, to 
coherent states yields
\begin{equation}
e_\lambda(f)\star e_\lambda(g) = e_\lambda(f + g+ fg). \label{2.3}
\end{equation}

As well as the $K$-transform, its dual operator $K^*$ will also play an 
essential role in our setting. In the sequel we denote by 
$\mathcal{M}_{\mathrm{fm}}^1(\Gamma)$ the set of all probability measures 
$\mu $ on $(\Gamma ,\mathcal{B}(\Gamma))$ with finite moments of all orders, 
i.e., 
\[
\int_\Gamma |\gamma _\Lambda |^n\,d\mu (\gamma )<\infty \quad 
\mathrm{for\,\,all}\,\,n\in \N\mathrm{\,\,and\,\,all}\,\,
\Lambda \in \mathcal{O}_c(\R^d).
\]
By the definition of a dual operator, given a 
$\mu\in \mathcal{M}_{\mathrm{fm}}^1(\Gamma)$, $K^*\mu =: \rho_\mu$ is a 
measure on $(\Gamma _0,\mathcal{B}(\Gamma _0))$ defined by 
\begin{equation}
\int_{\Gamma _0}G(\eta )\,d\rho _\mu(\eta )=\int_\Gamma \left(
KG\right) (\gamma )\,d\mu (\gamma ),  \label{Eq2.16}
\end{equation}  
for all $G\in B_{bs}(\Gamma_0)$. Following the terminology used in the 
Gibbsian case, we call $\rho_\mu$ the correlation 
measure corresponding to $\mu$. This definition shows, in particular, that 
$B_{bs}(\Gamma_0)\subset L^1(\Gamma_0,\rho_\mu)$. Moreover, on 
$B_{bs}(\Gamma_0)$ the inequality 
$\Vert KG\Vert_{L^1(\mu)}\leq \Vert G\Vert_{L^1(\rho_\mu)}$ holds, allowing
an extension of the $K$-transform to a bounded operator 
$K:L^1(\Gamma_0,\rho_\mu)\to L^1(\Gamma,\mu)$ in such a way that equality 
(\ref{Eq2.16}) still holds for any $G\in L^1(\Gamma_0,\rho_\mu)$. For the 
extended operator the explicit form (\ref{Eq2.9}) still holds, now $\mu$-a.e. 
This means, in particular,
$$
\left( Ke_\lambda (f)\right) (\gamma) = \prod_{x\in \gamma }(1+f(x)),\quad 
\mu \mathrm{-a.a.}\,\gamma\in\Gamma, 
$$
for all $\mathcal{B}(\R^d)$-measurable functions $f$ such that 
$e_\lambda(f)\in L^1(\Gamma_0,\rho_\mu)$.

All the notions described above as well as their relations are graphically 
summarized in the figure below. In the context of an infinite particle system 
this figure has a natural meaning. The state of such a system is described by 
a probability measure $\mu$ on $\Gamma$ and the functions $F$ on $\Gamma$ are 
considered as observables of the system and they represent physical 
quantities which can be measured. The measured values correspond to the 
expectation values $\int_\Gamma F(\gamma)\,d\mu(\gamma)$. In this context we 
call the functions $G$ on $\Gamma_0$ quasi-observables.
  


\vspace{1truecm}

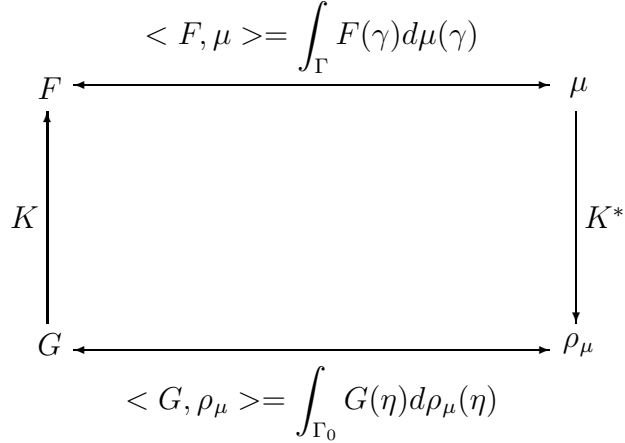
\begin{center}

\def\xpos {0}
\def\ypos {0}

\def\xlen {200}
\def\ylen {100}

\def\xinit {10}
\def\yinit {10}


\def\xlenline {180}           
\def\ylenline {80}           


\def\xposvectori {190}        
\def\yposvectori {90}        


\def\xletraposinit {-9}       
\def\xletraposend {191}       

\def\yletraposinit {-6}       
\def\yletraposend {91}       


\def\yexprtop {120}           
\def\yexprbottom {-18}        
\def\xexprleft {-14}          
\def\xexprright {203}         

\begin{picture} (\xlen, \ylen) (0, 0)


\put (\xinit,       \ypos) {\vector (1,0)  {\xlenline}}
\put (\xposvectori, \ypos) {\vector (-1,0) {\xlenline}}

\put (\xpos, \yinit)       {\vector (0,1) {\ylenline}}

\put (\xinit, \ylen)       {\vector (1,0) {\xlenline}}
\put (\xposvectori, \ylen) {\vector (-1,0) {\xlenline}}

\put (\xlen, \yposvectori) {\vector (0,-1) {\ylenline}}


\put (\xletraposinit, \yletraposinit) {\makebox(20,12)[t] {$G$}}
\put (\xletraposinit, \yletraposend)  {\makebox(20,12)[t]{$F$}}
\put (\xletraposend,  \yletraposend)  {\makebox(20,12)[t]{$\mu$}}
\put (\xletraposend,  \yletraposinit) {\makebox(20,12)[t]{$\rho _{_{}\mu}$}}


\put (\xpos, \yexprtop) 
{\makebox(\xlen, 12)[t] 
{$<F,\mu> = \displaystyle\int_{\Gamma} F(\gamma) d\mu (\gamma)$}}

\put (\xpos, \yexprbottom)
{\makebox(\xlen, 12)[t] {$<G,\rho _{_{}\mu}> = \displaystyle\int_{\Gamma_0} G(\eta) d\rho _{_{}\mu}(\eta)$}}

\put (\xexprleft, \ypos) 
{\makebox(12, \ylen)[l] {$K$}}

\put (\xexprright, \ypos) 
{\makebox(14, \ylen)[l] {$K^{*}$}}

\end{picture}

\end{center}

\vspace{1truecm}


\begin{example} On $\R^d$ consider the intensity measure $z\,dx$, $z > 0$, and 
the Poisson measure $\pi_z$ defined on $(\Gamma,\mathcal{B}(\Gamma))$ by 
\[
\int_\Gamma \exp \left( \sum_{x\in \gamma }\varphi (x)\right)\,d\pi_z(\gamma )
=\exp \left( z\int_{\R^d}\left( e^{\varphi (x)}-1\right)\,dx\right),\quad 
\varphi \in \mathcal{D}. 
\]
Here $\mathcal{D}:=C_0^\infty(\R^d)$ denotes the Schwartz space of all 
infinitely differentiable real-valued functions with compact support. The 
correlation measure corresponding to the Poisson measure $\pi_z$ is the 
so-called Lebesgue-Poisson measure 
$$
\lambda_z:=\sum_{n=0}^\infty \frac{z^n}{n!} m^{(n)},
$$
where each $m^{(n)}$, $n\in \N$, is the image measure on $\Gamma^{(n)}$ of 
the product measure $dx_1...dx_n$ under the mapping 
$\widetilde{(\R^d)^n}\ni (x_1,...,x_n)\mapsto\{x_1,...,x_n\}\in \Gamma^{(n)}$. For $n=0$ we set $m^{(0)}(\{\emptyset\}):=1$. This special case increases the 
importance of the coherent states and the space $B_{bs}(\Gamma_0)$ in our 
setting, mainly, due to the following two technical reasons, used throughout 
this work. First, $e_\lambda(f)\in L^p(\Gamma_0, \lambda_z)$ whenever 
$f\in L^p(\R^d, dx)$ for some $p\geq 1$, and, moreover,
\begin{equation}
\int_{\Gamma_0} \vert e_\lambda(f,\eta)\vert^p\,d\lambda_z(\eta)
= \exp \left( z\int_{\R^d} \vert f(x)\vert^p\,dx \right) . \label{4.1}
\end{equation}
Secondly, the space $B_{bs}(\Gamma_0)$ is dense in $L^2(\Gamma_0, \lambda_z)$.
\end{example}

\section{Gibbs measures on configuration spaces\label{Section2}}

Let $\phi:\R^d\to\R\cup\{+\infty\}$ be a pair potential, that is, a 
$\mathcal{B}(\R^d)$-measurable function such that $\phi(-x)=\phi(x)\in \R$ 
for all $x\in \R^d\setminus\{0\}$. For $\gamma\in\Gamma$ and 
$x\in \R^d\setminus\gamma$ we define a relative energy of interaction between 
a particle located at $x$ and the configuration $\gamma$ by 
\begin{eqnarray*}
E(x,\gamma ):=\left\{ 
\begin{array}{cl}
\displaystyle\sum_{y\in \gamma }\phi (x-y), & \mathrm{if\;}
\displaystyle\sum_{y\in \gamma }|\phi (x-y)|<\infty \\ 
&  \\ 
+\infty , & \mathrm{otherwise}
\end{array}
\right. .
\end{eqnarray*}
For $\gamma = \emptyset$ we set $E(x,\emptyset):= 0$. A grand 
canonical Gibbs measure (Gibbs measure for short) corresponding to a pair 
potential $\phi$ and an activity parameter $z>0$ is usually defined through 
the Dobrushin-Lanford-Ruelle equation. For convenience, we present here an 
equivalent definition through the Georgii-Nguyen-Zessin equation 
(\cite[Theorem 2]{NZ79}, see also \cite[Theorem 3.12]{KoKu98}, 
\cite[Appendix A.1]{K00}). More precisely, a probability measure $\mu$ on 
$(\Gamma ,\mathcal{B}(\Gamma))$ is called a Gibbs measure if it fulfills the 
integral equation
\begin{equation}
\int_\Gamma \sum_{x\in \gamma }H(x,\gamma )d\mu (\gamma )=z\int_\Gamma 
\int_{\R^d}H(x,\gamma \cup \{x\})e^{-E(x,\gamma )}\,
dxd\mu (\gamma ) \label{1.3}
\end{equation}
for all positive measurable functions $H:\R^d\times\Gamma\to \R$. In 
particular, for $\phi\equiv 0$, (\ref{1.3}) reduces to the Mecke identity, 
which yields an equivalent definition of the Poisson measure $\pi_z$ 
\cite[Theorem 3.1]{Me67}. For Gibbs measures, the corresponding 
correlation measures are always absolutely continuous with respect to the 
Lebesgue-Poisson measure $\lambda_z$. A Radon-Nikodym derivative 
$k_\mu:=\frac{d\rho_\mu}{d\lambda_z}$ is called the correlation function of 
the measure $\mu$. 

Throughout this work we shall consider potentials $\phi$ fulfilling the 
standard integrability (I) and stability (S) conditions: 
\begin{enumerate}
\item[\textbf{(I)}]$\qquad$  
$\displaystyle\int_{\R^d}\left| 1- e^{-\phi (x)}\right|\,dx<\infty$ . 
\item[\textbf{(S)}] There is a $B\geq 0$ such that 
$$
\forall\,\eta \in \Gamma _0,\quad 
E(\eta ):=\sum_{\{x,y\}\subset \eta }\phi (x-y)\geq -B|\eta |\quad
(E(\emptyset):=E(\{x\}):=0)
$$ 
\end{enumerate}
Let us note that if $\phi$ is semi-bounded from below, then condition (I) is 
equivalent to the integrability of $\phi$ on the set 
$\R^d\setminus\{\phi\geq 1\}$ whenever $\{\phi\geq 1\}$ has finite Lebesgue 
measure. Of course, the stability condition (S) implies the semi-boundeness 
of $\phi$ from below, namely, $\phi \geq -2B$ on $\R^d$. We will also use the 
superstability condition (SS), stronger than (S), and the lower regularity 
condition (LR), which may be found in \cite{Ru70}. 

For potentials fulfilling (I), (SS), and (LR), D.~Ruelle proved in 
\cite{Ru70} the existence of tempered Gibbs measures (Ruelle measures for 
short). For positive potentials, condition (I) is sufficient to 
insure the existence of Gibbs measures (see 
e.g.~\cite[Proposition 7.14]{KoKu98}, \cite[Proposition 2.7.15]{K00}). In 
either case, the corresponding correlation functions fulfil the so-called 
Ruelle bound (RB):
$$
\exists\ C>0:\  k_\mu(\eta)\leq e_\lambda(C,\eta)=C^{\vert \eta\vert},\quad 
\forall\,\eta\in \Gamma_0,
$$
cf.~\cite{Ru70}. Condition (RB) implies, in particular, that any Gibbs 
measure $\mu$ has all local moments finite, i.e., 
$\mu\in \mathcal{M}_{\mathrm{fm}}^1(\Gamma)$. 

\section{Infinitesimally invariant measures\label{Section3}}

In the recent work \cite{LK03}, the authors have shown that the operator $H$ 
defined on a proper set of cylinder functions by
$$
-(HF)(\gamma) := 
\sum_{x\in \gamma}\big(F(\gamma\setminus\{x\}) - F(\gamma)\big) 
+ z\int_{\R^d} e^{-E(x,\gamma)} \big(F(\gamma\cup\{x\}) - F(\gamma)\big)\,dx
$$
is the generator of an equilibrium Glauber-type dynamics. More precisely, for 
Gibbs measures $\mu$ corresponding to an activity parameter $z$ and a pair 
potential $\phi$ fulfilling either conditions (I), (SS), and (LR) or 
conditions $\phi\geq 0$ on $\R^d$ and (I), it is proved that $H$ is a 
positive definite 
symmetric operator on $L^2(\Gamma,\mu)$. This allows the use of standard 
Dirichlet forms techniques to construct a Markov process on $\Gamma$, called 
an equilibrium Glauber dynamics, having $\mu$ as an invariant measure. That 
is, $H^*\mu=0$ in the sense that
$$
\int_\Gamma (HF)(\gamma)\,d\mu(\gamma) = 0
$$
for all the cylinder functions $F$ as considered in \cite{LK03}. The dual 
relation between observables and states yields a further interpretation for 
this invariance result. Since the semigroup $T_t= e^{-tH}$ associated to $H$ 
on $L^2(\Gamma,\mu)$ is related to the Kolmogorov equation
\[
\frac{d}{dt}F_t=-HF_t,\quad t\geq 0  
\] 
on the space of observables, it is seen from the aforementioned dual relation 
that, for $\mu_t= T_t^*\mu = e^{-tH^*}\mu$, on the space of states one has
\begin{eqnarray*}
\left\{ 
\begin{array}{l}
\displaystyle\frac{d}{dt}\mu_t=-H^*\mu_t,\quad t\geq 0 \\ 
\\ 
\mu_0=\mu
\end{array}
\right. .  
\end{eqnarray*}
This consideration shows that the Gibbs measures studied in \cite{LK03} 
are stationary distributions of the dynamics generated by $H$ described 
above. One of our aims is to study the converse problem 
related to the question of whether all invariant measures are Gibbsian. As a 
first step for this purpose, we shall enlarged the class of invariant 
measures to a new class of measures outside of the semigroup setting. These 
new elements, so-called infinitesimally invariant measures, will be measures 
$\mu\in \mathcal{M}_{\mathrm{fm}}^1(\Gamma )$ such that, in a proper sense 
defined below (Definition \ref{Def1}), verify $H^*\mu=0$.

In order to define the notion of infinitesimally invariant measures corresponding to the 
operator $H$, first we shall extend the action of $H$ to the set of cylinder 
functions $\mathcal{FP}_{bc}(\Gamma)$. As 
$\mathcal{FP}_{bc}(\Gamma)=K\left (B_{bs}(\Gamma_0)\right )$, this procedure 
naturally leads to the operator $\hat H:= K^{-1}HK$ on the space of 
quasi-observables.

In the sequel we assume the potential $\phi$ to fulfil conditions (I) and 
(S). For functions $F\in \mathcal{FP}_{bc}(\Gamma)$, these assumptions are 
sufficient to insure that $HF$ is a well-defined function on $\Gamma$. This 
follows from the fact that for any $G\in B_{bs}(\Gamma_0)$ there are 
$\Lambda\in \mathcal{O}_c(\R^d), N\in\N_0$ and a $L\geq 0$ such that 
$G\!\!\upharpoonright _{\Gamma _0\backslash\left(\bigsqcup _{n=0}^N
\Gamma _\Lambda ^{(n)}\right)}\equiv 0$ and $\vert G\vert\leq L$, which implies that 
$F(\gamma):=\left (KG\right )(\gamma) = F\!\!\upharpoonright _{\Gamma _\Lambda }\!\!(\gamma_\Lambda)$ and 
$\vert F(\gamma)\vert\leq L(1+\vert \gamma_\Lambda\vert)^N$ for all 
$\gamma\in\Gamma$ (cf.~Section \ref{Section1}). Therefore,
$$
-(HF)(\gamma) = \sum_{x\in \gamma_\Lambda}\big(F(\gamma\setminus\{x\}) 
- F(\gamma)\big) + z\int_\Lambda e^{-E(x,\gamma)}
\big(F(\gamma\cup\{x\}) - F(\gamma)\big)\,dx,
$$
and the semi-boundeness of $\phi$ from below allows to majorize the integral 
by the function defined on $\Gamma$,
$$
e^{2B\vert\gamma\vert}\int_\Lambda 
\left(\left|F(\gamma_\Lambda\!\cup\!\{x\})\right| + \left|F(\gamma_\Lambda)\right|\right)\,dx \leq 
2Le^{2B\vert\gamma\vert}(2+\vert \gamma_\Lambda\vert)^Nm(\Lambda).
$$
Here, and below, $m(\Lambda)$ denotes the volume of a set $\Lambda$.

\begin{proposition}
\label{Prop3}The action of $\hat H$ on functions $G\in B_{bs}(\Gamma_0)$ is 
given by
$$
-(\hat{H}G)(\eta)= -\vert \eta \vert G(\eta) + z\int_{\R^d} \left(e_\lambda(e^{-\phi (x - \cdot)}-1)\star G(\cdot \cup \{x\})\right)(\eta)\,dx, 
$$
for all $\eta\in\Gamma_0$.
\end{proposition}

\noindent
\textbf{Proof.} According to the definitions of the operators $H$ and 
$\hat H$, for all $\eta\in \Gamma_0$ we find
\begin{eqnarray}
-(\hat{H}G)(\eta)&=& 
K^{-1}\Biggl(\sum_{x\in \cdot}\big((KG)(\cdot\setminus\{x\})- (KG)(\cdot)\big)\Biggr)(\eta) \nonumber \\
&&+ K^{-1}\Biggl(z\int_{\R^d} e^{-E(x,\cdot)}\big((KG)(\cdot\cup\{x\}) - (KG)(\cdot)\big)\,dx\Biggr)(\eta) \nonumber \\
&=&\sum_{\xi\subset\eta}(-1)^{\vert \eta\setminus\xi\vert}\sum_{x\in \xi}\big((KG)(\xi\!\setminus\!\{x\})- (KG)(\xi)\big) \label{2.1} \\
&&+z\sum_{\xi\subset\eta}(-1)^{\vert \eta\setminus\xi\vert}\!\int_{\R^d}\! e^{-E(x,\xi)}\big((KG)(\xi\cup\{x\}) - (KG)(\xi)\big)\,dx. \nonumber
\end{eqnarray}
A direct application of the definitions of the $K$-transform and its inverse 
mapping yields for the first sum in (\ref{2.1})
\begin{eqnarray*}
&&-\sum_{\xi\subset\eta}(-1)^{\vert \eta\setminus\xi\vert}\sum_{x\in \xi}\sum_{\rho\subset \xi\setminus\{x\}}G(\rho\cup\{x\})\\
&=&-\sum_{\xi\subset\eta}\sum_{x\in \xi}(-1)^{\vert \eta\setminus\xi\vert}(K(G(\cdot\cup\{x\})))(\xi\!\setminus\!\{x\})\\
&=&-\sum_{x\in \eta}\sum_{\xi\subset\eta\setminus\{x\}}(-1)^{\vert (\eta\setminus\{x\})\setminus\xi\vert}(K(G(\cdot\cup\{x\})))(\xi)\\
&=&-\sum_{x\in \eta}K^{-1}(KG(\cdot\cup\{x\}))(\eta\!\setminus\!\{x\})\\
&=&-\sum_{x\in \eta}G((\eta\!\setminus\!\{x\})\cup\{x\})= -\vert\eta\vert G(\eta).
\end{eqnarray*}
To compute the second sum in (\ref{2.1}), first observe that by the 
definition of the $K$-transform one has
\begin{eqnarray*}
&&\sum_{\xi\subset\eta}(-1)^{\vert \eta\setminus\xi\vert}\int_{\R^d} e^{-E(x,\xi)}\big((KG)(\xi\cup\{x\}) - (KG)(\xi)\big)\,dx\\
&=&\sum_{\xi\subset\eta}(-1)^{\vert \eta\setminus\xi\vert}\int_{\{x:x\not\in\xi\}} e^{-E(x,\xi)}\big((KG)(\xi\cup\{x\}) - (KG)(\xi)\big)\,dx\\
&=&\sum_{\xi\subset\eta}(-1)^{\vert \eta\setminus\xi\vert}\int_{\R^d} e^{-E(x,\xi)}\sum_{\rho\subset\xi}G(\rho\cup\{x\})\,dx\\
&=&\int_{\R^d}\sum_{\xi\subset\eta}(-1)^{\vert \eta\setminus\xi\vert}(Ke_\lambda(e^{-\phi(x-\cdot)}-1))(\xi)(K(G(\cdot\cup\{x\})))(\xi)\,dx.
\end{eqnarray*}
Therefore, by the action of the $K$-transform on the $\star$-convolution 
(\ref{1.5}), we finally obtain
\begin{eqnarray*}
&&\int_{\R^d}\sum_{\xi\subset\eta}(-1)^{\vert \eta\setminus\xi\vert}K\big(e_\lambda(e^{-\phi(x-\cdot)}-1)\star G(\cdot\cup\{x\})\big)(\xi)\,dx\\
&=&\int_{\R^d} K^{-1}\big(K(e_\lambda(e^{-\phi(x-\cdot)}-1)\star G(\cdot\cup\{x\}))\big)(\eta)\,dx\\
&=&\int_{\R^d} \big(e_\lambda(e^{-\phi(x-\cdot)}-1)\star G(\cdot\cup\{x\})\big)(\eta)\,dx.
\end{eqnarray*}
\hfill$\blacksquare \medskip$

\begin{proposition}
\label{Prop4}Given a $\mu\in\mathcal{M}_{\mathrm{fm}}^1(\Gamma)$ assume 
that the correlation measure $\rho_\mu$ is absolutely continuous with respect 
to the Lebesgue-Poisson measure $\lambda_z$, and the correlation function 
$k_\mu$ fulfills condition (RB) for some $C>0$. Then, 
$\hat H\left(B_{bs}(\Gamma_0)\right)\subset L^1(\Gamma_0,\rho_\mu)$. As a 
consequence, the operator $H$ maps the space $\mathcal{FP}_{bc}(\Gamma)$ into 
$L^1(\Gamma,\mu)$. 
\end{proposition}

\noindent
\textbf{Proof.} As any $G\in B_{bs}(\Gamma_0)$ fulfills $\vert G\vert\leq L$ 
and $G\!\!\upharpoonright _{\Gamma _0\backslash\left(\bigsqcup _{n=0}^N\Gamma _\Lambda ^{(n)}\right)}\equiv 0$ for some $L\geq 0, N\in\N_0$, and 
$\Lambda\in \mathcal{O}_c(\R^d)$, clearly one has
$$
\int_\Gamma \vert\eta\vert \vert G(\eta)\vert\,d\rho_\mu(\eta)\leq
NL\int_{\Gamma_\Lambda} C^{\vert\eta\vert}\,d\lambda_z(\eta)<\infty.
$$
Hence to prove the integrability of $\hat HG$ for $G\in B_{bs}(\Gamma_0)$ 
amounts to show the integrability of
$$
\int_{\R^d} \left(e_\lambda(e^{-\phi (x - \cdot)}-1)\star G(\cdot \cup \{x\})\right)(\eta)\,dx.
$$
In order to do this, first observe that any function $G\in B_{bs}(\Gamma_0)$ 
described as before verifies $\vert G\vert\leq Le_\lambda(1\!\!1_\Lambda)$, 
where $1\!\!1_\Lambda$ is the indicator function of $\Lambda$, and thus
\begin{eqnarray}
&&\int_{\Gamma_0}\left|\int_{\R^d} \left(e_\lambda(e^{-\phi (x - \cdot)}-1)\star G(\cdot \cup \{x\})\right)(\eta)\,dx\right|\,d\rho_\mu(\eta) \nonumber \\
&\leq& L\int_{\R^d}\int_{\Gamma_0}\left(e_\lambda\!\left(\left|e^{-\phi (x - \cdot)}-1\right|\right)\star e_\lambda(1\!\!1_\Lambda,\cdot \cup \{x\})\right)(\eta)\,d\rho_\mu(\eta)dx. \label{3.1}
\end{eqnarray}
The definition of the $\star$-convolution and its especially simple form 
(\ref{2.3}) for coherent states then allow rewriting the integrals in 
(\ref{3.1}) as
$$
\int_{\R^d}1\!\!1_\Lambda(x)\int_{\Gamma_0}
e_\lambda\!\left(1\!\!1_\Lambda + \left(1\!\!1_\Lambda +1\right) \left|e^{-\phi (x - \cdot)}-1\right|, \eta\right)\,d\rho_\mu(\eta)dx,
$$
which, due to the Ruelle boundeness, is bounded by
$$
\int_{\R^d}1\!\!1_\Lambda(x)\int_{\Gamma_0}
e_\lambda\!\left(C1\!\!1_\Lambda + C\left(1\!\!1_\Lambda +1\right) \left|e^{-\phi (x - \cdot)}-1\right|, \eta\right)\,d\lambda_z(\eta)dx.
$$
Assumption (I) combined with equality (\ref{4.1}) for the 
$\lambda_z$-expectation of a coherent state completes the proof showing that 
the latter expression may be bounded by
$$
m(\Lambda)\exp\left(zC\left(m(\Lambda) + 2\int_{\R^d}\left| e^{-\phi (x)}-1\right|\,dx\right)\right) < \infty.
$$
The last assertion arises from $K(\hat HG) = H(KG)$ for all 
$G\in B_{bs}(\Gamma_0)$, and the $K$-transform maps the space 
$L^1(\Gamma_0,\rho_\mu)$ into $L^1(\Gamma,\mu)$.\hfill$\blacksquare \medskip$

In this way Proposition \ref{Prop4} yields the following definition.

\begin{definition}
\label{Def1}A measure $\mu\in\mathcal{M}_{\mathrm{fm}}^1(\Gamma)$ as in 
Proposition \ref{Prop4} is called an infinitesimally invariant measure corresponding to 
$H$ whenever
$$
\int_\Gamma (HF)(\gamma)\,d\mu(\gamma) =0
$$
for all $F\in\mathcal{FP}_{bc}(\Gamma)$.
\end{definition}

\begin{theorem}
\label{Prop5}Let $\phi $ be a pair potential fulfilling (S), (I), and
(LR). Then any infinitesimally invariant measure corresponding to $H$ is Gibbsian. 
\end{theorem}

To prove this result we need the following lemma. We refer e.g.~to \cite{O02} 
for its proof.

\begin{lemma}
\label{Lmm2}Let $n\in \N$, $n\geq 2$, and $z>0$ be given. Then 
\begin{eqnarray*}
&&\int_{\Gamma _0}...\int_{\Gamma _0}G(\eta _1\cup ...\cup \eta_n)H(\eta
_1,...,\eta _n)d\lambda_z(\eta _1)...d\lambda_z(\eta _n)
\\
&=&\int_{\Gamma _0}G(\eta )\sum_{(\eta _1,...,\eta _n)\in \mathcal{P}_n(\eta
)}H(\eta _1,...,\eta _n)d\lambda_z(\eta )
\end{eqnarray*}
for all positive measurable functions $G:\Gamma _0\to\R$
and $H:\Gamma _0\times ...\times \Gamma _0\to\R$ with
respect to which both sides of the equality make sense. Here 
$\mathcal{P}_n(\eta )$ denotes the set of all partitions of $\eta $ in $n$ 
parts, which may be empty.
\end{lemma}

In particular, for $n = 3$, Lemma \ref{Lmm2} yields the 
following integration result for the $\star$-convolution.

\begin{lemma}
\label{Lmm3}For all positive measurable functions 
$H,G_1,G_2:\Gamma _0\to \R$ and all $z>0$ one has
\begin{eqnarray*}
&&\int_{\Gamma_0}H(\eta)(G_1\star G_2)(\eta)d\lambda_z(\eta) \\
&=&\int_{\Gamma_0}\!\int_{\Gamma_0}\!\int_{\Gamma_0}\!\!
H(\eta_1\cup\eta_2\cup\eta_3)
G_1(\eta_1\cup\eta_2)G_2(\eta_2\cup\eta_3)d\lambda_z(\eta_1)
d\lambda_z(\eta_2)d\lambda_z(\eta_3).
\end{eqnarray*}
\end{lemma}

\noindent
\textbf{Proof of Theorem \ref{Prop5}.} According to Proposition \ref{Prop4}, 
for any infinitesimally invariant measure $\mu$ corresponding to $H$ one has 
$$
\int_{\Gamma_0}(\hat HG)(\eta)k_\mu(\eta)\,d\lambda_z(\eta)=
\int_{\Gamma_0}(\hat HG)(\eta)\,d\rho_\mu(\eta)=
\int_\Gamma \left( H(KG)\right)(\gamma)\,d\mu(\gamma)=0
$$
for all functions $G\in B_{bs}(\Gamma_0)$. Concerning the first expectation, 
observe that an application of Lemma \ref{Lmm3} to the integral expression 
which appears in the definition of $\hat H$ yields
\begin{eqnarray*}
&&z\int_{\R^d} \int_{\Gamma_0}\big(e_\lambda(e^{-\phi (x - \cdot)}-1)\star 
G(\cdot \cup \{x\})\big)(\eta)k_\mu(\eta)d\lambda_z(\eta)dx\\
&=&z\int_{\R^d} \int_{\Gamma_0}\int_{\Gamma_0}\int_{\Gamma_0}
k_\mu(\eta_1\cup\eta_2\cup\eta_3)
G(\eta_1\cup \eta_2\cup \{x\})\cdot\\
&&\cdot \, e_\lambda(e^{-\phi (x - \cdot)}-1,\eta_2\cup\eta_3)
d\lambda_z(\eta_1)d\lambda_z(\eta_2)
d\lambda_z(\eta_3)dx\\
&=& \int_{\Gamma_0} d\lambda_z(\eta_3)
e_\lambda(e^{-\phi (x - \cdot)}-1,\eta_3)\\
&&\int_{\R^d} \int_{\Gamma_0}\int_{\Gamma_0}
k_\mu(\eta_1\cup\eta_2\cup\eta_3)
G(\eta_1\cup \eta_2\cup \{x\})\cdot \\
&&\cdot e_\lambda(e^{-\phi (x - \cdot)}-1,\eta_2)
zdxd\lambda_z(\eta_1)d\lambda_z(\eta_2)
\end{eqnarray*}
with
\begin{eqnarray*}
&&\int_{\R^d} \int_{\Gamma_0}\int_{\Gamma_0}
k_\mu(\eta_1\cup\eta_2\cup\eta_3)
G(\eta_1\cup \eta_2\cup \{x\})\cdot \\
&&\cdot e_\lambda(e^{-\phi (x - \cdot)}-1,\eta_2)
zdxd\lambda_z(\eta_1)d\lambda_z(\eta_2)\\
&=&\int_{\Gamma_0}G(\eta)\sum_{x\in \eta}
\sum_{\xi\subset \eta\setminus\{x\}} k_\mu((\eta\!\setminus\!\{x\})\cup\eta_3)
e_\lambda(e^{-\phi (x - \cdot)}-1,\xi)d\lambda_z(\eta),
\end{eqnarray*}
by an application of Lemma \ref{Lmm2} for $n=3$. Moreover, since
$$
\sum_{\xi\subset \eta\setminus\{x\}} e_\lambda(e^{-\phi (x - \cdot)}-1,\xi)
= e_\lambda(e^{-\phi (x - \cdot)},\eta\!\setminus\!\{x\}) 
= e^{-E(x,\eta\setminus\{x\})},
$$
we derive
\begin{eqnarray*}
&&\int_{\Gamma_0}G(\eta)\sum_{x\in \eta}
\sum_{\xi\subset \eta\setminus\{x\}} k_\mu((\eta\!\setminus\!\{x\})\cup\eta_3)
e_\lambda(e^{-\phi (x - \cdot)}-1,\xi)d\lambda_z(\eta)\\
&=&\int_{\Gamma_0}G(\eta)\sum_{x\in \eta}
e^{-E(x,\eta\setminus\{x\})}k_\mu((\eta\!\setminus\!\{x\})\cup\eta_3)
d\lambda_z(\eta).
\end{eqnarray*}
As a result
\begin{eqnarray*}
&&z\int_{\R^d} \int_{\Gamma_0}\big(e_\lambda(e^{-\phi (x - \cdot)}-1)\star 
G(\cdot \cup \{x\})\big)(\eta)k_\mu(\eta)d\lambda_z(\eta)dx\\
&=&\int_{\Gamma_0}G(\eta)\sum_{x\in \eta}
e^{-E(x,\eta\setminus\{x\})}\cdot \\
&&\cdot\int_{\Gamma_0}e_\lambda(e^{-\phi (x - \cdot)}-1,\rho)
k_\mu((\eta\!\setminus\!\{x\})\cup\rho)\,d\lambda_z(\rho)d\lambda_z(\eta).
\end{eqnarray*}
In this way for all $G\in B_{bs}(\Gamma_0)$ one finds
\begin{eqnarray*}
0&=&\int_{\Gamma_0} (\hat HG)(\eta)k_\mu(\eta)\,d\lambda_z(\eta)\\
&=&\int_{\Gamma_0}\vert \eta \vert G(\eta)k_\mu(\eta)d\lambda_z(\eta) 
- \int_{\Gamma_0}G(\eta)\sum_{x\in \eta}e^{-E(x,\eta\setminus\{x\})}\\
&&\cdot\int_{\Gamma_0}e_\lambda(e^{-\phi (x - \cdot)}-1,\rho)k_\mu((\eta\!\setminus\!\{x\})\cup\rho)\,d\lambda_z(\rho)d\lambda_z(\eta).
\end{eqnarray*}
This implies 
\begin{equation}
\vert \eta \vert k_\mu(\eta) = \sum_{x\in \eta}e^{-E(x,\eta\setminus\{x\})}
\int_{\Gamma_0}e_\lambda(e^{-\phi (x - \cdot)}-1,\rho)
k_\mu((\eta\!\setminus\!\{x\})\cup\rho)\,d\lambda_z(\rho) \label{6.1}
\end{equation}
for $\lambda_z$-a.a.~$\eta \in \Gamma_0$. Note that in terms of the adjoint 
operator ${\hat H}^*$ of $\hat H$ on $L^2(\Gamma_0,\lambda_z)$, equality 
(\ref{6.1}) means ${\hat H}^*k_\mu=0$. We proceed to show the equivalence 
between equation (\ref{6.1}) and the so-called Kirkwood-Salsburg equation 
((KS)-equation for short), i.e., 
\[
k_\mu(\eta\!\cup\!\{x\})=e^{-E(x,\eta)}\!\int_{\Gamma_0}\!\!
e_\lambda (e^{-\phi (x - \cdot )}-1,\rho )k_\mu (\eta\cup \rho )\,
d\lambda_z(\rho ),\quad \lambda_z\!\otimes\!dx\!-\!a.e.
\]
Once this is proved, the proof then naturally follows by Proposition 
\ref{8Prop8.2.2} below, due to \cite{Ru70}. For non-translation invariant 
potentials, a similar result has been proved in \cite[Section 2.6]{K00}. 

Let $k$ be a correlation function solving the (KS)-equation. Then 
$$
\!\!e^{-E(x,\eta\setminus\{x\})}
\!\!\!\int_{\Gamma_0}\!\!\!e_\lambda(e^{-\phi (x - \cdot)}-1,\rho)
k((\eta\setminus\{x\})\cup\rho)\,d\lambda_z(\rho)
=k(\{x\}\cup(\eta\setminus\{x\})) = k(\eta),
$$
and summing both sides for all $x\in\eta$ yields equation (\ref{6.1}). To 
check the converse implication, let us first rewrite equation (\ref{6.1}) in 
the simpler form 
$$
\sum_{x\in\eta}I(x,\eta\!\setminus\!\{x\}) = 0,
$$
where
\begin{eqnarray*}
&&I(x,\eta\!\setminus\!\{x\}) :=\\
&&\!\!k(\{x\}\!\cup\!(\eta\!\setminus\!\{x\}))\!
-\!e^{-E(x,\eta\setminus\{x\})}\!
\int_{\Gamma_0}\!\!\!e_\lambda(e^{-\phi (x - \cdot)}-1,\rho) 
k((\eta\!\setminus\!\{x\})\cup\rho)d\lambda_z(\rho).
\end{eqnarray*}
A straighforward application of Lemma \ref{Lmm2} for $n=2$ then yields 
\begin{eqnarray*}
0&=&\int_{\Gamma_0}G(\eta)\sum_{x\in\eta}I(x,\eta\!\setminus\!\{x\})\,
d\lambda_z(\eta)\\
&=&\int_{\Gamma_0}\int_{\R^d}G(\eta\cup\{x\})I(x,\eta)\,zdxd\lambda_z(\eta),
\end{eqnarray*}
for all functions $G\in B_{bs}(\Gamma_0)$. This implies that 
$dx\!\otimes\!\lambda_z$-a.e.~$I(x,\eta) = 0$, that is, the 
(KS)-equation.\hfill$\blacksquare \medskip$

\begin{proposition}
\label{8Prop8.2.2}Let $\phi $ be a pair potential fulfilling (S), (I), and
(LR). Given a $\mu \in \mathcal{M}_{\mathrm{fm}}^1(\Gamma )$ assume that the
correlation measure $\rho_\mu$ is absolutely continuous with respect to a
Lebesgue-Poisson measure $\lambda_z$ for some $z>0$, and the correlation 
function $k_\mu$ fulfills (RB). Then, $\mu$ is a Gibbs measure corresponding 
to $\phi $ and the activity $z$ if and only if $k_\mu$ solves the 
(KS)-equation. 
\end{proposition}

\begin{remark}

Calculations similar to those in the proof of Theorem \ref{Prop5} show 
that, in terms of Bogoliubov functionals $L_\mu$ corresponding to 
infinitesimally invariant measures $\mu$,
$$
L_\mu(\varphi) :=\int_\Gamma \prod_{x\in \gamma }(1+\varphi(x))\,d\mu(\gamma),
\quad \varphi\in\mathcal{D},
$$
one finds the equality
$$
\int_{\R^d}\varphi(x)
\left (L_\mu((\varphi + 1)(e^{-\phi (x - \cdot)}-1)+ \varphi)-
\frac{\delta L_\mu(\varphi)}{\delta \varphi(x)}\right )\,zdx=0,
$$
for all $\varphi\in\mathcal{D}$. Here 
$\frac{\delta L_\mu(\varphi)}{\delta \varphi(x)}$ denotes the first 
variational derivative of $L_\mu$ at $\varphi$. This leads to the well-known 
equilibrium Bogoliubov equation introduced in \cite{Bog46}
$$
\frac{\delta L_\mu(\varphi )}{\delta \varphi (x)}=L_\mu\left( (1+\varphi )\left(
e^{-\phi (x -\cdot )}-1\right) +\varphi \right), \quad 
dx \mathrm{-\mathit{a.e.}},
$$
which yields an equivalent description of Gibbs measures (see \cite{KoKu99c}, 
\cite{KoKuOl02}, \cite{K00}, and also \cite{Naz85}).   
\end{remark}

\addcontentsline{toc}{section}{References}
\bibliographystyle{alpha}
\bibliography{/home/oliveira/artigos/Oliveira}

\begin{thebibliography}{BRW02}

\bibitem[BCC02]{BCC02}
L.~Bertini, N.~Cancrini, and F.~Cesi.
\newblock The spectral gap for a {G}lauber-type dynamics in a continuous gas.
\newblock {\em Ann. Inst. H. Poincar\'e Probab. Statist.}, 38:91--108, 2002.

\bibitem[Bog46]{Bog46}
N.~N. Bogoliubov.
\newblock {\em Problems of a Dynamical Theory in Statistical Physics}.
\newblock Gostekhisdat, Moscow, 1946.
\newblock (in Russian). English translation in J. de Boer and G. E. Uhlenbeck
  (editors), {\it Studies in Statistical Mechanics\/}, volume 1, pages 1--118.
  North-Holland, Amsterdam, 1962.

\bibitem[BRW02]{BRW02}
V.~I. Bogachev, M.~R{\"o}ckner, and F.-Y. Wang.
\newblock Invariance implies {G}ibbsian: some new results.
\newblock BiBoS Preprint n.~02-12-106, University of Bielefeld, 2002.

\bibitem[Dob94]{Do94}
R.~L. Dobrushin.
\newblock A mathematical approach to foundations of statistical mechanics.
\newblock Preprint ESI, Vienna, 1994.

\bibitem[FLO97]{FLO97}
J.~Fritz, C.~Liverani, and S.~Olla.
\newblock Reversibility in infinite {H}amiltonian systems with conservative
  noise.
\newblock {\em Comm. Math. Phys.}, 189(2):481--496, 1997.

\bibitem[Fri82]{Fr82}
J.~Fritz.
\newblock Stationary measures of stochastic gradient systems, infinite lattice
  models.
\newblock {\em Z. Wahrsch. verw. Gebiete}, 59:479--490, 1982.

\bibitem[Fri86]{Fr86}
J.~Fritz.
\newblock On the stationary measures of anharmonic systems in the presence of a
  small thermal noise.
\newblock {\em J. Statist. Phys.}, 44(1--2):25--47, 1986.

\bibitem[FRZ98]{FRZ98}
J.~Fritz, S.~Roelly, and H.~Zessin.
\newblock Stationary states of interacting {B}rownian motions.
\newblock {\em Studia Sci. Math. Hungar.}, 34(1--3):151--164, 1998.

\bibitem[HS81]{HS81}
R.~Holley and D.~W. Stroock.
\newblock Diffusions on an infinite-dimensional torus.
\newblock {\em J. Funct. Anal.}, 42(1):29--63, 1981.

\bibitem[KK02]{KoKu99}
{Yu}.~G. Kondratiev and T.~Kuna.
\newblock Harmonic analysis on configuration space {I}. {G}eneral theory.
\newblock {\em Infinite Dimensional Analysis, Quantum Probabilities and Related
  Topics}, 5(2):201--233, 2002.

\bibitem[KK03a]{KoKu98}
{Yu}.~G. Kondratiev and T.~Kuna.
\newblock Correlation functionals for {G}ibbs measures and {R}uelle bounds.
\newblock {\em Methods of Functional Analysis and Topology}, 9(1):9--58, 2003.

\bibitem[KK03b]{KoKu99c}
{Yu}.~G. Kondratiev and T.~Kuna.
\newblock Harmonic analysis on configuration space {II}. {B}ogoliubov
  functional and equilibrium states.
\newblock {I}n preparation, 2003.

\bibitem[KKO02]{KoKuOl00b}
{Yu}.~G. Kondratiev, T.~Kuna, and M.~J. Oliveira.
\newblock On the relations between {P}oissonian white noise analysis and
  harmonic analysis on configuration spaces.
\newblock BiBoS Preprint n.~02-02-072, University of Bielefeld, 2002.

\bibitem[KKO03]{KoKuOl02}
{Yu}.~G. Kondratiev, T.~Kuna, and M.~J. Oliveira.
\newblock Non-equilibrium stochastic dynamics of continuous systems and
  {B}ogoliubov functionals.
\newblock In preparation, 2003.

\bibitem[KL03]{LK03}
{Yu}. Kondratiev and E.~Lytvynov.
\newblock Glauber dynamics of continuous particle systems.
\newblock math.{PR}/0306252 at ar{X}iv.org, 2003.

\bibitem[Kun99]{K00}
T.~Kuna.
\newblock {\em Studies in Configuration Space Analysis and Applications}.
\newblock PhD thesis, Bonner Mathematische Schriften Nr.~324, University of
  Bonn, 1999.

\bibitem[Mec67]{Me67}
J.~Mecke.
\newblock Station\"are zuf\"allige {M}a\ss e auf lokalkompakten {A}belschen
  {G}ruppen.
\newblock {\em Z. Wahrsch. verw. Gebiete}, 9:36--58, 1967.

\bibitem[Naz85]{Naz85}
G.~I. Nazin.
\newblock Method of the generating functional.
\newblock {\em J.~Sov.~Math.}, 31:2859--2886, 1985.

\bibitem[NZ79]{NZ79}
X.~X. Nguyen and H.~Zessin.
\newblock Integral and differential characterizations of the {G}ibbs process.
\newblock {\em Math.~Nachr.}, 88:105--115, 1979.

\bibitem[Oli02]{O02}
M.~J. Oliveira.
\newblock {\em Configuration Space Analysis and Poissonian White Noise
  Analysis}.
\newblock PhD thesis, Faculty of Sciences, University of Lisbon, 2002.

\bibitem[Rue70]{Ru70}
D.~Ruelle.
\newblock Superstable interactions in classical statistical mechanics.
\newblock {\em Comm. Math. Phys.}, 18:127--159, 1970.

\end{thebibliography}

\end{document}